\documentclass[a4paper,twocolumn,showpacs,preprintnumbers,amsmath,amssymb,aps,nofootinbib]{revtex4}
\usepackage{epsfig}
\usepackage{graphicx,amsmath}
\usepackage{graphicx}
\usepackage{dcolumn}
\usepackage{bm}

\newcommand{\lbfig}[1]{\refstepcounter{fig} \label{#1} }
\newcounter{fig}

\newcounter{One}
\newcounter{Two}
\newcounter{Three}
\def\putunder#1#2{\mathrel{
\setbox0=\hbox{#1}\setbox1=\hbox{\scriptsize #2}
\dimen0=-0.5\wd0 \advance\dimen0 by -0.5\wd1
\dimen1=0.5\wd0 \advance\dimen1 by -0.5\wd1
\hbox{\box0\kern\dimen0%
\vbox to 0pt {\hbox{\lower 0.7em \box1}\vss}%
\kern\dimen1}
}}
\def\beq{\begin{equation}}
\def\eeq{\end{equation}}
\def\bea{\begin{eqnarray}}
\def\eea{\end{eqnarray}}
\newcommand{\gsim}{\lower.7ex\hbox{$\;\stackrel{\textstyle>}{\sim}\;$}}
\newcommand{\lsim}{\lower.7ex\hbox{$\;\stackrel{\textstyle<}{\sim}\;$}}

\begin{document}

\preprint{DESY 05-066}

\title{Analytically derived limits on short-range fifth forces
\\[0.3cm]from quantum states of neutrons in the Earth's gravitational field}


\author{A. Westphal}
\altaffiliation[Corresponding author. ]{\\ Present address: SISSA \& INFN, Via Beirut
2-4, I-34014 Trieste\\ Email:
westphal@sissa.it} \affiliation{Deutsches Elektronen-Synchrotron, Notkestra\ss e 85, D-22607 Hamburg, Germany}
\author{H. Abele}
\affiliation{%
Physikalisches Institut der Universit\"{a}t Heidelberg\\
Philosophenweg 12\\
69120 Heidelberg, Germany
}%

\author{S. Bae\ss ler}%
\affiliation{%
Institut f\"ur Physik, Universit\"{a}t Mainz\\
Staudinger Weg 7\\
55128 Mainz, Germany
}%

\date{March 9, 2007}

\begin{abstract}

Recently, quantum states of ultra-cold neutrons in the Earth's
gravitational field have been observed for the first time. From the
fact that they are consistent with Newtonian gravity on the
10\%-level, analytical limits on $\alpha$ and $\lambda$ of
short-range Yukawa-like additional interactions are derived between
$\lambda$ = 1 $\mu m$ and 1 $mm$. We arrive for $\lambda\geq 10\,\mu
m$ at $\alpha < 2\cdot 10^{11}$ at 90\% confidence level. This
translates into a limit $g_s g_p/\hbar c<2\cdot 10^{-15}$ on the
pseudo-scalar coupling of axions in the previously experimentally
unaccessible astrophysical axion window.
\end{abstract}


\maketitle

\section{Introduction}

A gravitational bound quantum system has been realized
experimentally. In this experiment, ultra-cold neutrons (UCN) are
confined in between a bottom mirror and the gravitational potential
of the Earth~\cite{nes1,nes2,nes2005}. The neutrons are slow enough
that they are reflected from the mirror at all angles of incidence.
Therefore, the mirror can be modeled by an infinite high potential
step. The neutrons are found in discrete quantum states of the
gravity potential. Between the UCN source and a UCN detector one
places a quantum state absorber at a certain hight above the mirror.
\begin{figure}[ht]
\begin{center}
\epsfig{file=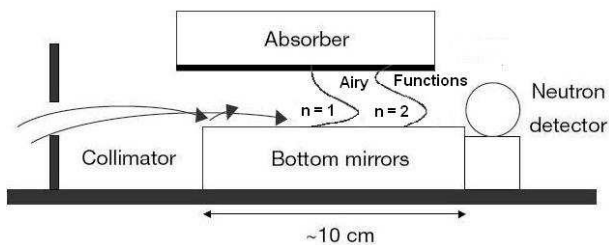,scale=.4} \lbfig{fig.1}
\caption{\small
\newline Experimental setup: Neutrons pass through the mirror-absorber system. The Airy functions for the first two bound quantum states are shown. The absorber is at variable height.}
\end{center}
\end{figure}
No neutrons except those in sufficiently low quantum states as given
by the absorber height can pass through the slit between the mirror
and absorber, and higher, unwanted states are removed and scattered
out of the experiment (see Fig.~\ref{fig.1}).

A side-effect of this experiment is its sensitivity to additional
short ranged forces at length scales below 10~$\mu$m~\cite{ab,nes3},
while all electromagnetic effects are extremely suppressed compared
to gravity~\cite{ab}. The quantum states probe Newtonian gravity
between $10^{-9}$ and $10^{-3}$~m and the experiment places limits
for gravity-like forces there. So far, significant limits on
hypothetical forces from this experiment are mainly obtained from
measurements of the ground state or from a fit to the
data~\cite{ab,nes3}.  Other experimental limits on extra forces are
derived from mechanical experiments and can be found, e.g.,
in~\cite{Bordag,ad,Chia,Fischb2,Kapner}. In the light of recent
theoretical developments in higher dimensional field
theory~\cite{Arkani1,Arkani2,Ant2,Antoniadis,dim}, gauge fields
could mediate forces that are $10^{6}$ to $10^{12}$ times stronger
than gravity at sub-millimeter distances (depending on the size of
the extra dimensions), exactly in the interesting range of this
experiment and might give a signal in an improved setup. Recent
theoretical developments support this original idea of strong forces
with bulk gauge fields. Burgess et al.~\cite{Bu} predict deviations
from Newton's law on the micron scale on the basis of supersymmetric
large extra dimensions (SLED). The basic idea behind this proposal
is to modify gravity at small distances in such a way as to explain
the smallness of the observed cosmological constant. A radius $R$ of
10 microns as well as the necessary interaction strength may turn
out to be well-motivated. We presents limits for additional
interactions using an elementary particle, the neutron. The limits
are derived in regions of the parameter space, where these
interactions can be treated perturbatively.

\section{Additional short range forces}

We begin with recalling the standard parametrization of a fifth
force by means of a Yukawa potential describing the low energy limit
of the exchange of massive particles, proceeding then to the action
of a perturbatively weak Yukawa-like fifth force on the
gravitational bound states in different regimes of its range
$\lambda$. On the assumption that the form of the non-Newtonian
potential is given by the Yukawa expression, for a source mass $m$
and distance $r$ the modified Newtonian potential $\phi(r)$ has the
form \beq \phi(r) = -G_4\frac{m}{r}(1+\alpha\cdot{e^{-r/\lambda}}),
\eeq where $\lambda$ is the range over which the corresponding force
acts and $\alpha$ is the strength normalized relative to Newtonian
gravity. $G_4$ is the gravitational constant. The mass of this
extended source will modify the Newtonian potential if strong
non-Newtonian forces are present, and this can be seen if neutrons
are present. For small distances $z$ from the mirror, say several
micrometers, we consider the mirror as an infinite half-space with
mass density $\rho$. By replacing the source mass $m$ by $\int\rho
dV$, the Yukawa-modification of the potential $\phi(r)$ has the form
\beq \Delta\phi(z,\lambda) = -2\pi\cdot{\rho}\alpha\lambda^2
G_4\cdot{e^{-z/\lambda}}. \label{dpot}\eeq

Thus, the effective gravitational potential close above the mirror
(close means: heights $z<<D$, $D$: diameter of the mirror) is given
as: \beq \phi(z)=\underbrace{g\cdot
z}_{\phi_0}\underbrace{-2\pi\cdot\alpha\cdot\lambda^2\cdot
G_4\cdot\rho\cdot e^{-z/\lambda}}_{\Delta\phi}\label{pot}, \eeq
where $\rho$ denotes the mass density of the mirror material (glass
in our experiment).

The absorber has (besides its Fermi pseudopotential) in presence of
a fifth force a Yukawa-like additional potential attached to it just
like the bottom mirror. Then $\Delta\phi$ becomes \beq
\Delta\phi(z)=-2\pi\cdot\alpha\cdot\lambda^2\cdot
G_4\cdot\left(\rho_1\cdot e^{-z/\lambda}+\rho_2\cdot
e^{-(h-z)/\lambda}\right)\label{pot2} \eeq where $h$ denotes as
before the absorber height and $\rho_1,\rho_2$ the mass densities of
the bottom mirror and the absorber, respectively. In some parts of
this paper, we can ignore the absorber.

Within an unchanged Newtonian gravitational potential \beq
\phi_0(z)=g\cdot z\label{newt} \eeq the bound eigenstates of the UCN
are given by Airy functions $Ai(z)$. This function behaves similar
to a sine wave below $h_0^{(0)}$ and approaches zero exponentially
above this classical turning point height. With \bea
\label{turning}h_n^{(0)}&=&R\cdot
\left[\frac{3\pi}{2}\cdot\left(n+\frac{3}{4}\right)\right]^{2/3}\;\eea
\beq R=\left(\frac{\hbar^2}{2m_n^2 g} \right)^{1/3}
\;,\;\;\nonumber\\C_n=\left\langle
n_g^{(0)}\right.\left|n_g^{(0)}\right\rangle^{-1/2}\label{hn} \eeq
one finds \bea \left|n_g^{(0)}\right\rangle&=&\int
dz\cdot\psi_{n,g}^{(0)}(z)\left|z\right\rangle\;\;,\;\;\;\left\langle
z'\right.\left|z\right\rangle=2\pi\cdot\delta(z'-z)\nonumber\\
\psi_{n,g}^{(0)}(z)&=&C_n\cdot
Ai((z-h_n^{(0)})/R)\;,\;C_n=:\frac{\tilde{C}_n}{\sqrt{R}}\label{eigenf}\\
&=&C_n\cdot \frac{d
Ai}{dz}(-h_n^{(0)}/R)\cdot\frac{z}{R}+{\cal O}((z/R)^2)\nonumber\\
&\simeq&C_n\cdot
\frac{1}{2}\left(\frac{h_n^{(0)}}{R}\right)^{1/4}\cdot\frac{z}{R}+{\cal
O}((z/R)^2)\label{eigens} \eea Here, $h_n^{(0)}$ denotes the
position of the last turning point of the bound state
$\psi_{n,g}^{(0)}(z)$, which coincides with the turning point height
of the classical motion of a particle with energy
$E_n^{(0)}=mgh_n^{(0)}$. $h_n^{(0)}$ has been determined
~\cite{nes3} for the first two states to be \bea
\label{expturning}h_0^{\rm
exp}&=12.2\pm 0.7_{\rm stat}\pm 1.8_{\rm syst}\\
h_1^{\rm exp}&=21.6\pm 0.7_{\rm stat}\pm 2.2_{\rm syst}. \eea

With this analogy in mind, $h_n^{0}$ is considered as the height of
the wave function. The ground state is opaque for neutrons as long
as it sufficiently overlaps with the absorber above the mirror. This
is not the case when the absorber position exceeds the height
$h_n^{(0)}$ following the exponentially decaying tail of the bound
state's Airy function, and neutrons in that state are transmitted to
the detector, see Fig. 1. In the WKB approximation the
$\psi_{n,g}^{(0)}$ are: \bea
\psi_{n,g}^{(0)}(z)&\simeq&C_n\cdot\frac{1}{2}\cdot\left(\frac{h_n^{(0)}-z}{R}\right)^{-1/4}
\nonumber\\
&\phantom{\simeq}&\cdot\sin\left\{\frac{1}{R}\cdot\int_0^zdu
\sqrt{\frac{1}{R}\left[h_m^{(0)}-u\right]}\right\}
\;\;\;.\label{WKB} \eea
\section{Our method}
We start from the observation that at macroscopic absorber heights
of several microns, where light is easily transmitted, no neutrons
pass through the gap between the mirror and the
absorber~\cite{nes2005}. A neutron in state $n$ is only transmitted
as long the absorber height is higher than $h_n^{(0)}$.

Taking additional gravity-like forces into account, first order
perturbation theory predicts a shift of the energy eigenvalue of the
ground state. The effect of the Yukawa term in eq.~(\ref{pot}) in
first order is\beq \Delta E_n^{(1)}=m\cdot\left\langle
n_g^{(0)}\right|\Delta\phi(z)
\left|n_g^{(0)}\right\rangle\;\;\;.\label{dE} \eeq This energy shift
is accompanied by a shift of the turning point $h_n^{0}$ by some
additional height $\delta h$, in this way changing the onset of the
transmission in the experiment. Our measurement follows the
Newtonian expectation.

\def\Hc{\hbox{$h_{\rm c}$}}
The absorber consists of glass material, where the surface has an
approximate gaussian roughness of 0.75 $\mu$m. The height of the
absorber has been calibrated with wire-spacers of known thickness, a
mechanical comparator, and a long focus microscope. An absolute
height calibration of better than 0.5 $\mu$m has been achieved.
Next, we recall that the absorber forms a hard wall for the very
slow neutrons. This results in a squeezing of the bound
states~\cite{Wes06,wes} and a small energy shift, which is
considered in the following way: The squeezing increases the bound
state energy eigenvalue by an amount $\Delta E_{\rm squeez}=m\cdot
g\cdot \Delta_{\mathrm{Sq}}$, where $\Delta_{\mathrm{Sq}}\approx
0.8\,\mu m$ denotes the corresponding shift of the turning point of
the squeezed wave function. As a result, the turning point of the
squeezed bound states in presence of a real absorber is given by
$h_n=h_n^{(0)}+\Delta_{\mathrm{Sq}}$. We now calculate the influence
of a hypothetical fifth force on the turning points of the first two
states (Eq.\ref{turning}) and compare them with the experimental
results given in Eq.\ref{expturning}. The actual neutron
transmission of the $n^{\rm th}$ bound state is observed at a mean
absorber height $h_n+\Delta_T$. $\Delta_T$ has been chosen to be
3$\mu m$, since 3$\mu m$ above the turning point, the slope of the
transmission curve shows that state $n$ is transmitted. Summarizing
both effects, a bound state $n$ transmits neutrons, which are
clearly visible in the detector if the absorber is at a height
$h_n+\Delta_{\mathrm{T}}=h_n^{(0)}+\Delta_{\mathrm{T}}+\Delta_{Sq}$.

In this paper, we derive limits on 5th forces, which are largely
independent from a precise knowledge of the absorber height and
unaffected by the features of the absorber. The limits are obtained
from the fact that the ground state and the first excited state wave
functions are differently affected by non-Newtonian forces. The fifth force would introduce the height shift $\delta h$, so that we measure $h=h_n^{(0)}+\Delta_{\mathrm{T}}+\Delta_{Sq}+\delta h$. In
particular, the difference of the turning point heights of the
ground state and the first excited state \beq \Delta
h^{(0)}:=h_0^{(0)}-h_1^{(0)}\approx 10.3\,\mu m\label{Dh}\eeq is
consistent with the measurement $\Delta h^{\rm exp}=9.4\pm 1.2\,\mu
m$. This error includes in addition to the statistical uncertainty a
0.5$\mu$m calibration uncertainty~\cite{nes2005} and a 30\%
uncertainty due to model dependence. (N.B.: Only the relative error
of a height measurement enters.) Since this relative quantity is
mostly insensitive to absolute offsets of the measurement process of
$h$, we will use it to derive limits on the presence of additional
short-range fifth forces, i.e., we can exclude a fifth force induced
shift \beq |\delta\Delta h|
> 1.64\cdot\sigma_{\Delta h}=2.0\,\mu m\label{expdat}\eeq at 90\% confidence level.
In the case of Eq.~(\ref{limit3}), the choice of $\Delta_{\mathrm{T}}$ and
$\Delta_{\mathrm{Sq}}$ enters, but our limits are largely independent
from these quantities.

To apply this method it is essential to derive the connection
between the perturbative correction to the energy eigenvalue of a
given bound state, which is induced by the additional force, and a
possible accompanying shift $\delta h_n$ of the turning point of the
$n^{\rm th}$ state's wave function. The stationary Schr\"odinger equation for the bound states
\beq \frac{\partial^2}{\partial
z^2}\psi_{n,g}(z)=-\frac{2
m}{\hbar^2}\cdot\left[E_n-m\cdot\phi(z)\right]\psi_{n,g}(z),\label{schrod}
\eeq implies a turning point condition given by \bea
\left.\frac{\partial^2\psi_{n,g}(z)}{\partial z^2}\right|_{z=h_n}&=&0\nonumber\\
&=&-\frac{2
m_n}{\hbar^2}\cdot\left[E_n-m\cdot\phi(h_n)\right]\psi_{n,g}(h_n)\nonumber\\
&\Rightarrow&E_n-m\cdot\phi(z)=0\textrm{ for }z=h_n\;.\label{turn}
\eea Write now for the height of the wave function under the
presence of the additional force $h_n=h_n^{(0)}+\delta
h_n+\Delta_{\mathrm{Sq}}$, $E_n=E_n^{(0)}+\Delta E_n^{(1)}+\Delta
E_{\rm squeez}$ and use the fact that $E_n^{(0)}=mg\cdot h_n^{(0)}$.
Then one arrives at \beq \left.\Delta
E_n^{(1)}\right|_{h=h_n+\Delta_{\mathrm{T}}}-m\cdot\left[g\cdot\delta
h_n+\Delta\phi(h_n^{(0)}+\delta
h+\Delta_{\mathrm{Sq}})\right]=0\label{turn2} \eeq with
$\Delta\phi(z)$ given by eq.~\eqref{pot}. Note, that the piece
linear in $\Delta_{\mathrm{Sq}}$ from the squeezing of the bound
states between absorber and mirror cancels against the corresponding
energy shift $\Delta E_{\rm squeez}$. It is immediately clear from
here that concerning the situation with just a bottom mirror for
$\lambda << h_n^{(0)}$ one has $\Delta\phi(h_n^{(0)})\approx 0$ and
eq.~\eqref{turn2} simplifies in this regime to \beq \delta
h_n=\frac{1}{g}\left\langle n_g^{(0)}\right|\Delta\phi(z)
\left|n_g^{(0)}\right\rangle\;\;\;.\label{corr} \eeq In the opposite
case $\lambda \gsim h_n^{(0)}$ one may linearize $\Delta\phi(z)$ in
$z/\lambda$, which approximation then has to be used simultaneously
in exploiting eq.~\eqref{turn2} and eq.~\eqref{dE} to calculate
$\Delta E_n^{(1)}$. 

For later use, let us note one further property
of the above turning point condition. $\Delta\phi(z)$ may contain a
constant, position-independent part (for instance, the above linear
approximation in the case $\lambda \gsim h_n^{(0)}$ generically
produces such a constant piece). Now, from considering the general
behaviour of the exact solution it is clear, that changing the
potential by an arbitrary constant must leave the whole bound state
as well as its turning point unchanged though it does change the
energy eigenvalue of the state. This fact is clearly contained in
eq.~\eqref{turn2}: Imagine adding a constant $\Delta\phi_{const}$ to
$\Delta\phi(z)$. Then its contribution to $\Delta E_n^{(1)}$ is
given by $\Delta\phi_{const}$ and thus cancels out against the same
term in $\Delta\phi(z)$. Thus one may expand $\Delta\phi(z)$ around
any given convenient point $z$ and drop the constant piece.

In either case we proceed then by extracting the turning point
shifts $\delta h_{0,1}$ from eq.~\eqref{turn2} for the ground state
and the first excited state, respectively. Forming the difference
\beq \delta\Delta h=\delta h_0-\delta h_1 \label{dDh} \eeq allows us
then to extract limits on the strength $\alpha$ of the additional
fifth force as a function of its range $\lambda$ by demanding eq.~\eqref{expdat}, the
experimental constraint at 90\% C.L.

\section{Bottom mirror and no absorber - Case I: small $\lambda \ll h_0^{(0)}$}


Consider now the first case $\lambda<<h_0^{(0)}$. Here, for a
perfect absorber as said above, eq.~\eqref{corr} provides a good
description of the shift of the ground state turning point (see
the general discussion of Sect. III). Then the linear
approximation of the bound states given in Sect. II suffices to
calculate eq.~\eqref{corr} since $\Delta\phi$ then is confined to
a region $\simeq\lambda<<h_0^{(0)}$. Since it is the ground state
which defines the non-penetration region and thus that feature of
the measured neutron transmission function which responds most
sensitively to energy or equivalently $h_n$-shifts, we evaluate
eq.~(\ref{corr}) for $n=0$: \bea \delta
h_n&=&\frac{1}{g}\left\langle
n_g^{(0)}\right|\Delta\phi \left|n_g^{(0)}\right\rangle\nonumber\\
&\simeq&-\tilde{C}_n^2\cdot\frac{\pi\cdot\alpha\cdot\lambda^2\cdot
G_4\cdot\rho}{2\cdot g \cdot R^3}\nonumber\\
&\phantom{\simeq}&\cdot\sqrt{\frac{h_n^{(0)}}{R}}\int_0^{\infty}dz\cdot
z^2 e^{-z/\lambda} \nonumber\\
&\simeq&-\tilde{C}_n^2\sqrt{\frac{h_n^{(0)}}{R}}\cdot\frac{\pi\cdot\alpha\cdot\lambda^5\cdot
G_4\cdot\rho}{g \cdot R^3} \;\;\;\;.\label{shift} \eea Forming
$\delta\Delta h$ according to eq.~\eqref{dDh} we demand the
experimental constraint eq.~\eqref{expdat}. Therefore we arrive at
an exclusion limit in $\alpha$-$\lambda$-space given by \beq
|\alpha|\leq \frac{1}{|\tilde{C}_0^2\sqrt{\frac{h_0^{(0)}}{R}}-
\tilde{C}_1^2\sqrt{\frac{h_1^{(0)}}{R}}|}\cdot\frac{g \cdot
R^3}{\pi\cdot\lambda^5\cdot G_4 \cdot\rho}\cdot\delta\Delta h
\sim\lambda^{-5}\label{limit} \eeq which behaves symmetrical for
attractive and repulsive forces.


\section{Bottom mirror and no absorber - Case II: large $\lambda >> h_0^{(0)}$}

In the second case $\lambda>>h_0^{(0)}$. Then $\Delta\phi\simeq
const.$ over the whole range where $\psi_{n,g}^{(0)}$ is sizable.
However, since constant pieces of the potential drop out from the
turning point condition one has to apply it now in its precise
form carefully expanding the fifth force potential to linear order
in $z$ and $h_0$, respectively. This yields
\[\Delta\phi(z)=2\pi\cdot\alpha\cdot\lambda^2\cdot
G_4\cdot\rho\cdot\left(1-\frac{z}{\lambda}\right)+{\cal
O}\left(z^2/\lambda^2\right)\;\;.\] Instead of $\Delta\phi$ we
plug its contribution linear in $z$ into eq.~\eqref{dE} (recall
that constant pieces of $\Delta\phi$ later will drop out of the
turning point condition anyway) \bea \Delta
E_n^{(1)}&=&m\cdot\left\langle
n_g^{(0)}\right|\Delta\phi \left|n_g^{(0)}\right\rangle\nonumber\\
&\simeq&-C_n^2\cdot m\cdot2\pi\cdot\alpha\cdot\lambda^2\cdot
G_4\cdot\rho\nonumber\\
&\phantom{\simeq}&\cdot \int_0^{\infty}dz\cdot
\left|\psi_{n,g}^{(0)}(z)\right|^2
(-z/\lambda)\nonumber\\
&\simeq&m\cdot2\pi\cdot\alpha\cdot\lambda\cdot
G_4\cdot\rho\cdot\langle z\rangle_n\label{shift2} \eea where
eq.~\eqref{turn2} allows to compute the turning point shift
$\delta h$ of the ground state given by \bea
2\pi\cdot\alpha\cdot\lambda\cdot
G_4\cdot\rho\left(\langle z\rangle_n-(h_n^{(0)}+\Delta_{\mathrm{Sq}})\right)&\phantom{=}&\nonumber\\
-\left(g+2\pi\cdot\alpha\cdot\lambda\cdot
G_4\cdot\rho\right)\cdot\delta h_n&=&0\;\;.\nonumber \eea From this
follows - via forming the quantity $\delta\Delta h$ again - a limit
in the case of a repulsive interaction ($\alpha<0$) given by \beq
\alpha\geq-\frac{g}{2\pi\cdot
G_4\cdot\rho}\cdot\frac{1}{1+\frac{\Delta h^{(0)}-(\langle
z\rangle_0-\langle z\rangle_1)}{\delta\Delta
h}}\cdot\frac{1}{\lambda}\sim\lambda^{-1}\;\;\;.\label{limit2rep}
\eeq For the attractive case ($\alpha>0$) a smooth solution of
eq.~\eqref{turn2} exists for all $0\geq\delta\Delta h>-(\Delta
h^{(0)}-(\langle z\rangle_0-\langle z\rangle_1))$ which yields a
limit \beq \alpha\leq\frac{g}{2\pi\cdot
G_4\cdot\rho}\cdot\frac{1}{\frac{\Delta h^{(0)}-(\langle
z\rangle_0-\langle z\rangle_1)}{|\delta\Delta
h|}-1}\cdot\frac{1}{\lambda}\sim\lambda^{-1}\;\;\;.\label{limit2att}
\eeq Here it is $\langle z\rangle_0\approx 1.56\cdot R\approx
9.15\,\mu m$ and $\langle z\rangle_1\approx 2.73\cdot R\approx
16.0\,\mu m$.


The transition between the two regimes of small and large
$\lambda$ takes place just around $\lambda\approx 5\ldots 7\,\mu
m$ as it can be seen by comparing eq.s~\eqref{limit} and
\eqref{limit2rep}.

\section{Bottom mirror and real absorber - Large $\lambda$\label{sec:largelambda}}

A real absorber now has (besides its Fermi pseudopotential) in
presence of a fifth force a Yukawa-like additional potential
attached to it just like the bottom mirror. Then $\Delta\phi$
becomes \beq \Delta\phi(z)=-2\pi\cdot\alpha\cdot\lambda^2\cdot
G_4\cdot\left(\rho_1\cdot e^{-z/\lambda}+\rho_2\cdot
e^{-(h-z)/\lambda}\right)\label{pot2} \eeq where $h$ denotes as
before the absorber height and $\rho_1,\rho_2$ the mass densities
of the bottom mirror and the absorber, respectively. For large
$\lambda$ it makes sense to expand eq.~(\ref{pot2}) in
$(z-h/2)/\lambda$ around $z=h/2$: \bea
\Delta\phi(z)&\simeq&-2\pi\cdot\alpha\cdot\lambda^2\cdot G_4\cdot
e^{-h/2\lambda}\cdot\bigg(\rho_1+\rho_2\nonumber\\&
&{}-(\rho_1-\rho_2)\cdot \frac{z-h/2}{\lambda}+\frac{(\rho_1
+\rho_2)\cdot(z-h/2)^2}{2\cdot\lambda^2}\nonumber\\& &{}+{\cal
O}(\frac{z^3}{\lambda^3})\bigg) \nonumber\\&=&const.-\pi\,\alpha\,
G_4\cdot
e^{-h/2\lambda}\cdot\big[-2\lambda\cdot(\rho_1-\rho_2)\cdot
z\nonumber\\& &{} +(\rho_1+\rho_2)\cdot(z^2-z\cdot
h)\big]+\lambda^2\,{\cal O}(\frac{z^3}{\lambda^3})\label{pot2b}\;.
\eea

\subsection{Special case $\rho_1=\rho_2$}

Now in our case both mirror and absorber are made from glass, so
$\rho_1=\rho_2=\rho$. Then the direct linear term in the former
expansion vanishes - and thus all terms containing positive powers
of $\lambda$! For large $\lambda$ thus just the quadratic term
remains since it is independent of $\lambda$, all higher terms are
suppressed by negative powers of $\lambda$ and vanish in this
limit. As an immediate consequence this means that for large
$\lambda$ the presence of an absorber with a density equal to that
of the mirror leads to a $\lambda$-independent limit on the
strength $\alpha$ of an additional Yukawa-like interaction.

To determine this limit one computes the energy correction
eq.~\eqref{dE} induced by the potential eq.~\eqref{pot2b} now to
first order in perturbation theory \beq \Delta
E_n^{(1)}=-m\cdot2\pi\alpha\cdot G_4\cdot\rho\cdot
e^{-h/2\lambda}\cdot\left\langle z^2-z\cdot
h\right\rangle_n\;,\label{corr3b} \eeq where $\langle
z^2-zh\rangle_n=\langle z^2-zh_n^{(0)}\rangle_n-\langle
z\rangle_n\,(\Delta_{\mathrm{T}}+\Delta_{\mathrm{Sq}}+\delta h_n)$
for $h=h_n^{(0)}+\delta
h_n+\Delta_{\mathrm{T}}+\Delta_{\mathrm{Sq}}$ denotes the
expectation value of the z-dependent part of the correction (for the
z-independent part dropping out of the turning point condition
eq.~\eqref{turn2} see the general discussion above) to the potential
with respect to the ground state $\psi_{0,g}^{(0)}$. Then one
inserts eq.~\eqref{pot2b} into the turning point condition
eq.~\eqref{turn2} carefully evaluating to linear order in $\delta
h_n$, uses the value of the turning point $h_n=h_n^{(0)}+\delta
h_n+\Delta_{\mathrm{Sq}}$ in the experiment, and obtains \bea
\left.\Delta
E_n^{(1)}\right|_{h=h_n+\Delta_{\mathrm{T}}}\hspace*{-0.8ex}&-&
m\cdot g\cdot\delta h_n\label{turnp3}\\
&-&\hspace*{-3.2ex}\underbrace{m\cdot\Delta\phi\left(h_n^{(0)}+\delta
h_n+\Delta_{\mathrm{Sq}}\right)}_{m\cdot2\pi\alpha\cdot
G_4\cdot\rho\cdot
e^{-h/2\lambda}\cdot\Delta_{\mathrm{T}}\,(h_n^{(0)}+\delta
h_n+\Delta_{\mathrm{Sq}})}\hspace{-2ex}=\hspace{1ex}0\;\;.\nonumber \eea Determining $\delta
h_n$ from these equations for $n=0,1$  allows us to form
subsequently $\delta\Delta h=\delta h_0-\delta h_1$. Demanding then
$-1.64\sigma_{\Delta h}<\delta\Delta h<1.64\sigma_{\Delta h}$ yields
then a limit on the interaction strength $\alpha$ given by
\begin{widetext} \bea |\alpha|&\leq&g\,|\delta\Delta h| \cdot
\Bigg|\Bigg\{\pi \left(e^{-\frac{h_0}{2 \lambda }} \left[\langle
z^2\rangle_0+\Delta _{\mathrm{T}} (h_0^{(0)}+\Delta
_{\mathrm{Sq}})-\langle z\rangle_0 (h_0^{(0)}+\Delta
_{\mathrm{T}}+\Delta
_{\mathrm{Sq}})\right]\right.\nonumber\\&&\left.-e^{-\frac{h_1}{2 \lambda }}
\left[\langle z^2\rangle_1+\Delta _{\mathrm{T}} (h_1^{(0)}+\Delta
_{\mathrm{Sq}})-
\langle z\rangle_1 (h_1^{(0)}+\Delta _{\mathrm{T}}+\Delta _{\mathrm{Sq}})\right]\right.\nonumber\\
&& \left.+\delta\Delta h\,\left(e^{-\frac{h_0}{2 \lambda }} (\Delta _{\mathrm{T}}-\langle
z\rangle_0)+e^{-\frac{h_1}{2 \lambda }} (\Delta _{\mathrm{T}}-\langle z\rangle_1)\right) + \sqrt{D}\right) \rho
\,G_4\Bigg\}^{-1}\Bigg|\label{limit3}\\
&&\nonumber\\ &&\nonumber \\
D&=&\left\{e^{-\frac{h_0}{2 \lambda }} \left[\langle
z^2\rangle_0+\Delta _{\mathrm{T}} (h_0^{(0)}+\Delta
_{\mathrm{Sq}})-\langle z\rangle_0 (h_0^{(0)}+\Delta
_{\mathrm{T}}+\Delta _{\mathrm{Sq}})\right]-e^{-\frac{h_1}{2 \lambda
}} \left[\langle
z^2\rangle_1+\Delta _{\mathrm{T}} (h_1^{(0)}+\Delta _{\mathrm{Sq}})\right.\right.\nonumber\\
&&\left.\left.-\langle z\rangle_1 (h_1^{(0)}+\Delta
_{\mathrm{T}}+\Delta _{\mathrm{Sq}})\right]+e^{-\frac{h_0}{2 \lambda
}} (\Delta _{\mathrm{T}}-\langle z\rangle_0) \delta\Delta h+e^{-\frac{h_1}{2
\lambda }}
(\Delta _{\mathrm{T}}-\langle z\rangle_1) \delta\Delta h\right\}^2\nonumber\\
&&-4 e^{-\frac{h_0+h_1}{2 \lambda }} \delta\Delta h \left\{(\Delta _{\mathrm{T}}-\langle z\rangle_1) \left[\langle z^2\rangle_0+\Delta
_{\mathrm{T}} (h_0^{(0)}+\Delta
_{\mathrm{Sq}})-\langle z\rangle_0 (h_0^{(0)}+\Delta _{\mathrm{T}}+\Delta _{\mathrm{Sq}})\right.\right.\nonumber\\
&&\left.\left.+(\Delta _{\mathrm{T}}-\langle z\rangle_0)
\delta\Delta h\right]-(\Delta _{\mathrm{T}}-\langle z\rangle_0)
\left[\langle z^2\rangle_1+\Delta _{\mathrm{T}} (h_1^{(0)}+\Delta
_{\mathrm{Sq}})-\langle z\rangle_1 (h_1^{(0)}+\Delta _{\mathrm{T}}+\Delta
_{\mathrm{Sq}})\right]\right\}\;\;,\nonumber \eea
\end{widetext} which is plotted in Fig.~\ref{fig.2} These limits become essentially independent of $\lambda$ for
$\lambda > h\approx h_1^{(0)}$. The result for the repulsive
case contains a pole, which in the worst case, e.g.for $\lambda$ =
500 $\mu $m is at \bea\delta\Delta h\approx 2.2\,\mu
m\;\;,\label{pole}\eea and signals a breakdown of the perturbative
approach. The pole in the attractive case above corresponds to the
fact that for an attractive Yukawa potential at the mirror there is
a $|\delta h|$, above which the wave function begins to get sucked
into the Yukawa potential. However, once the attractive potential
gets strong enough for this to happen, the perturbative expansion
around the original states ceases to be a good approximation, which
explains such a pole: If one added the higher orders of perturbation
theory to the above constraint, the above pole would occur at just
that $\delta h$, where the corresponding wave function becomes
non-perturbatively deformed, which would then coincide with validity
boundary of perturbation theory derived below in eq.~\eqref{limit4}.

Finally, for really large $\lambda$ the constant term in the
potential eq.~\eqref{pot2b} eventually becomes energetically
dominant. Since this term in any case has to stay significantly
smaller than the total kinetic energy $E_{UCN}$ of the ultra-cold
neutrons (velocity of $\sim 5\,m/s$) entering the waveguide
(otherwise they cease to be transmitted entirely, which is not
observed), the condition
\[\Delta\phi(z)\simeq-2\pi\cdot\alpha\cdot\lambda^2\cdot
G_4\cdot\rho\lsim 0.1\cdot E_{UCN}\] limits in the case of an
attractive interaction the validity of eq.~\eqref{limit3} to
$\lambda\lsim 10\,mm$. For the case of a repulsive interaction
this leads to a further decrease of the bound of the strength
$\alpha$ above $\lambda\sim 10\,mm$ of the 5th force like \beq
|\alpha|\leq 3.8\cdot
10^{12}\cdot\left(\frac{10\,mm}{\lambda}\right)^2\;\;.\label{limit3b}\eeq

\subsection{General case $\rho_1\neq\rho_2$}

Let us now discuss what happens if $\rho_1\neq\rho_2$. In this
case the linear term in eq.~\eqref{pot2b} reappears. Since this
term is $\sim\lambda$ it will dominate the quadratic term
considered so far for large $\lambda$. Thus, if $\rho_1-\rho_2$ is
not tuned to be very small, the limit on the Yukawa interaction is
again given through calculating $\delta\Delta h$ by the
eq.s~\eqref{limit2rep} and~\eqref{limit2att}, however, with $\rho$
replaced by $\rho_1-\rho_2$ and $\Delta\phi$ given by the piece
linear in $z$ of eq.~\eqref{pot2b}
\bea\left|\alpha\right|&\lsim&\frac{g}{2\pi\cdot
G_4\cdot(\rho_1-\rho_2)}\cdot\frac{1}{\frac{\Delta h^{(0)}
-(\langle z\rangle_0-\langle z\rangle_1)}{\delta\Delta
h}+1}\cdot\frac{e^{\bar{h}/2\lambda}}{\lambda}\nonumber\\
&&\label{limit2b}\\
&\sim&\lambda^{-1}\;,\;\;\lambda> \bar{h}\;\;.\nonumber\eea Here we have defined $\bar{h}=(h_1+h_0)/2\approx 18.9\,\mu$ and neglected factors $\exp(\pm\Delta h/2\lambda)$ which are suppressed for $\lambda>\Delta h=-\Delta h_0/2\approx 5.2\,\mu$. This treatment is perturbative.

We can cross-check this result by looking again at the potential
eq.~\eqref{pot2b}, and add its linear piece to the earth's
gravitational potential $\phi_0=g\cdot z$\[\phi(z)=g\cdot
z+2\pi\cdot\alpha\cdot\lambda\cdot G_4\cdot(\rho_1-\rho_2)\cdot
z\;\;.\]Clearly, the case $\rho_1-\rho_2$ amounts to a
renormalization of $g$
given by \bea g\rightarrow g^{\prime}&=&g+\delta g\label{dg}\\
&\phantom{=}&\delta g=2\pi\cdot\alpha\cdot\lambda\cdot
G_4\cdot(\rho_1-\rho_2)\;\;.\nonumber\eea Since the turning points
of the bound state Airy functiosn are given by eq.~\eqref{hn} the
turning point shifts in this situation, where $g$ renormalizes,
using eq.~\eqref{dg} write as \bea \delta h_n&=&
\left[\frac{3\pi}{2}\cdot\left(n+\frac{3}{4}\right)\right]^{2/3}
\cdot\frac{\partial R}{\partial g}\cdot\delta
g\nonumber\\
&=&-\left[\frac{3\pi}{2}\cdot\left(n+\frac{3}{4}\right)\right]^{2/3}
\cdot R\cdot\frac{\delta g}{3g}\nonumber\\
&=&-h_n^{(0)}\cdot\frac{2\pi\cdot\alpha\cdot\lambda\cdot
G_4\cdot(\rho_1-\rho_2)}{3\cdot g}\;\;.\label{dhreng}\eea This
relation then implies after forming $\delta\Delta_h=\delta
h_0-\delta h_1$ through the experimental constraint
eq.~\eqref{expdat} $\left|\delta\Delta h\right|<1.6\,\mu m$ a
symmetrical limit on the strength of the Yukawa interaction \beq
\left|\alpha\right|\leq \frac{3\cdot g}{2\pi\cdot
G_4\cdot(\rho_1-\rho_2)}\cdot\frac{\left|\delta\Delta
h\right|}{\Delta
h^{(0)}}\cdot\frac{1}{\lambda}\sim\lambda^{-1}\label{limit6}\eeq
which is the same functional dependence as given in
eq.~\eqref{limit2b} ($\Delta h^{(0)}-(\langle z\rangle_0-\langle
z\rangle_n)\approx \Delta h^{(0)}/3$) for large $\lambda>\bar{h}$.

Let us shortly note now, that this limit for large $\lambda>20\,\mu
m$ can be relatively easily converted into bounds of the strength of
the matter couplings of axions. Axion interactions with a range
within $20\,\mu m<\lambda<200\,mm$ (corresponding to axion masses
$10^{-6}\,\textrm{eV}<m_a<10^{-2}\,\textrm{eV}$), the 'axion
window', are still allowed by the otherwise stringent constraints
posed by cosmological data (see e.g.~\cite{kolb,rose}). They lead to
a potential which is proportional to the 5th force potential
eq.~\eqref{dpot}, however, the axion-induced potential changes sign
with the direction of the neutron spin polarization relative to the
mirror. Thus, the relevant limit is again given by
eq.~\eqref{limit6} with, however, one small but significant change
induced by this pseudoscalar nature of the axion interaction,
namely, that $\rho_2\to-\rho_2$ (see \cite{wilcz}). This, in turn,
implies that the bound on the scalar-pseudoscalar axion interaction
could be derived using just eq.~\eqref{limit6}, however, with
$\rho_2$ replaced as $\rho_2\to-\rho_2$ (or the perturbative limit
eq.~\eqref{limit2b}, with again the replacement $\rho_2\to-\rho_2$).

\subsection{$\rho_2\to-\rho_2$ - axion limits in the astrophysical axion window}

To make the last statement more precise, note that an axion would
feel a CP-violating spin-dependent interaction in presence of
matter given by \cite{wilcz} \beq V(\vec{r})=\hbar g_s g_p
\frac{\vec{\sigma}\cdot\vec{n}}{8\pi m c}\left(\frac{1}{\lambda
r}+\frac{1}{r^2}\right)\,e^{-r/\lambda}\;\;.\label{axpot1}\eeq
Here, $\vec{\sigma}$ denotes the neutron spin and $\vec{n}$ is a
unit vector presumably related to the geometry of the macroscopic
matter configuration. Integrating this potential over the geometry
of our mirror-absorber-system gives a Yukawa-potential
contribution of the form~\cite{ValAx} \bea
\Delta\phi(z)&=&-\alpha_a\cdot\frac{\hbar^2\rho_1\lambda}{8
m^3}\,e^{-z/\lambda}+\alpha_a\cdot\frac{\hbar^2\rho_2\lambda}{8
m^3}\,e^{-(h-z)/\lambda}\nonumber\\&&\\ &=&-2\pi\alpha_{\rm
eff.}\cdot\lambda^2\cdot G_4\cdot(\rho_1\,
e^{-z/\lambda}-\rho_2\,e^{-(h-z)/\lambda})\;\;,\label{axpot2}\nonumber\eea
where we have used that \beq\alpha_{\rm
eff.}=\alpha_a\cdot\frac{\hbar^2}{16\pi G_4\cdot
m^3}\cdot\lambda^{-1}\;\;,\;\alpha_a:=\frac{g_s g_p}{\hbar
c}\;\;.\label{alphaax}\eeq Note, how the switch in the sign of the
two terms in eq.~\eqref{axpot2} arises from the fact that
$\vec{\sigma}\cdot\vec{n}/|\vec{\sigma}\cdot\vec{n}|=+1$ for
$\vec{n}$ the unit normal on the mirror but
$\vec{\sigma}\cdot\vec{n}/|\vec{\sigma}\cdot\vec{n}|=-1$ for
$\vec{n}$ the unit normal on the absorber. Plugging now
eq.~\eqref{alphaax} into eq.~\eqref{limit2b} we arrive at a limit
for the dimensionless axion coupling strength $\alpha_a$ given by
\beq\left|\alpha_a\right|\lsim\frac{4\,m^3
g}{\hbar^2\,\rho}\cdot\frac{1}{\frac{\Delta h^{(0)} -(\langle
z\rangle_0-\langle z\rangle_1)}{\delta\Delta h}+1}\cdot
e^{\bar{h}/2\lambda}\label{limitaxion}\eeq where we used that
$\rho_1=\rho_2=\rho$ in our case. For $\lambda \gsim\bar{h}$ this
leads at 90\% confidence level to a $\lambda$-independent upper
limit on the axion interaction strength \beq \frac{g_s g_p}{\hbar
c}\lsim 2\cdot 10^{-15}\;\;.\label{axionlimit2}\eeq

\section{Bottom mirror and real absorber - Small $\lambda$}

This situation is similar to that of Sect. VI, except for two
crucial differences: Firstly, due to $\lambda\ll h$ we have to use full potential of eq.~\eqref{pot2} instead of its expansion in eq.~\eqref{pot2b}. Secondly, the presence of the absorber at
$h=h_n+\Delta_{\mathrm{T}}$, the turning point, yields a relatively large
contribution to $\Delta\phi(h_n)$ since there the exponential
factor in the Yukawa potential is
$e^{(h_n-h)/\lambda}=e^{-\Delta_{\mathrm{T}}/\lambda}$. This inserted into
eq.~\eqref{turn2} and using eq.~\eqref{shift} to compute $\Delta
E_n^{(1)}$ yields an equation for $\delta h$ given by \bea
\Delta E_n^{(1)}&-&m\cdot g\cdot\delta h_n\nonumber\\
&-&\underbrace{m\cdot\Delta\phi_{Abs.}(h_n)}_{= -m\cdot2\pi\cdot
G_4\cdot\rho\cdot\alpha\cdot\lambda^2\cdot e^{-\Delta_{\mathrm{T}}/\lambda}}=0\nonumber\\
&\hspace{-3.ex}\textrm{with: }&\Delta
E_n^{(1)}=-\tilde{C}_n^2\,\sqrt{\frac{h_n^{(0)}}{R}}\cdot
\pi\cdot
m\,\alpha\,G_4\,\rho\cdot\frac{\lambda^5}{R^3}\nonumber \eea Here it has
been made use of the exponentially decaying tail of a bound
state Airy function for $z>h_n$ that suppresses the contribution
of the absorber attached Yukawa potential to $\Delta E_n^{(1)}$.
Therefore we neglect the absorber for the calculation of the
energy shift. Calculating for the first two states $\delta\Delta h=\delta h_0-\delta h_1$
and reshuffling of this equation to extract $\alpha$ then writes
as a limit symmetrical in its sign which reads exactly the same as
the simple one of eq.~\eqref{limit} (!) \bea
|\alpha|&\leq&\frac{1}{|\tilde{C}_0^2\sqrt{\frac{h_0^{(0)}}{R}}-
\tilde{C}_1^2\sqrt{\frac{h_1^{(0)}}{R}}|}\cdot\frac{g \cdot
R^3\cdot \delta\Delta h}{\pi\cdot\lambda^5\cdot G_4 \cdot\rho}\label{limit5}\\
&\sim&\lambda^{-5}\;\;\textrm{for small }\lambda,\lambda\lsim
5\,\mu m\;\;.\nonumber \eea The fact that the limit is symmetrical rests on neglecting the absorber in the calculation of the energy shift. Were this taken into account it would render the limit asymmetrical. Thus, this limit is shown for the attractive case in
Fig.~\ref{fig.2} (thick red long-dashed line) for
$\lambda\geq1.2\,\mu m$. For $\lambda<1.2\,\mu m$ this limit
violates the general perturbativity bound discussed in the next
Section. As we expect the limit - as before - for the repulsive case to be weaker than for the attractive case, it would be completely outside the general perturbativity bound for $\lambda<h$, which is why we do not display it.

\begin{widetext}
\bea \Delta h_0^{(2)}&=&\frac{1}{m_n\cdot
g}\sum_{m>0}\frac{\left|\left\langle
m_g^{(0)}\right|m_n\cdot\Delta\phi\left|0_g^{(0)}\right\rangle\right|^2}{\underbrace{m_n\cdot
g\cdot(h_0^{(0)}}_{=E_0^{(0)}}-h_m^{(0)})}\nonumber\\
&\simeq&-\sum_{m>0}C_0^2C_m^2\cdot\frac{\pi^2\cdot\alpha^2\cdot\lambda^4\cdot
G_4^2\cdot\rho^2}{2\cdot g^2 \cdot
R^3\cdot\sqrt{h_0^{(0)}h_m^{(0)}}}\cdot\left[\int_0^{\infty}dz\cdot
\sin(\phi_0(z))\sin(\phi_m(z))\cdot
e^{-z/\lambda}\right]^2\nonumber\\&\phantom{=}&\nonumber\\&\phantom{=}&\textrm{with:}\;\;\;C_m\simeq
\frac{e^{\kappa_0}}{\sqrt{R}}\cdot
m^{-1/3}\;,\;\;\kappa_0\simeq0.71\textrm{ and
}\;\phi_m(z)=\frac{1}{R}\cdot\int_0^zdu\sqrt{\frac{1}{R}\left[h_m^{(0)}-u\right]}
\;\;\;(\textrm{use eq.~(\ref{WKB}) around }z=0\;)\nonumber\\&\phantom{=}&\nonumber\\
&\simeq&48^{1/3}\frac{2\cdot\pi^2\cdot\alpha^2\cdot\lambda^{10}\cdot
G_4^2\cdot\rho^2}{g^2 \cdot
R^7}\cdot\underbrace{\sum_{m>0}\frac{1}{\left[(m+3/4)^{2/3}\cdot\frac{\lambda^2}{R^2}+1\right]^4}
\cdot\frac{(1+3/4m)^{1/3}}{(m+3/4)^{2/3}-(3/4)^{2/3}}}_{=:\;\sigma(\lambda)\;\simeq\;
\zeta(3)\;:\;\;convergent}\;\;\;\;.\label{shift4} \eea
\end{widetext}

\section{Validity of perturbation theory}

There remains now the question of the range of validity of
perturbation theory. The limit of eq.~(\ref{limit}) for small
$\lambda$ in principle might lead to values of $\alpha$ and
$\lambda$ where $\Delta\phi<E_0/m$ does not hold any more at least
for $z=0$. To estimate the validity of perturbation theory we will
resort to comparing the $2^{nd}$ order perturbation theory to the
$1^{st}$ order (see e.g.~\cite{schwab}). Then $\Delta
E_0^{(2)}<\Delta E_0^{(1)}$ will provide us with a validity limit
of perturbation theory. The $2^{nd}$ order perturbation theory is
given as seen in eq.~(\ref{shift4}). The sum $\sigma(\lambda)$ is
convergent and can be evaluated numerically via integral
approximations. This, in turn, then yields the validity range of
perturbation theory for small  and large $\lambda$, respectively,
as: \beq
\alpha\leq\left\{\begin{array}{ll}\frac{3^{2/3}\cdot\pi^{4/3}\cdot
g\cdot R^4}{4\cdot\pi^2\cdot 48^{1/3}\cdot
G_4\cdot\rho}\cdot\frac{1}{\lambda^5}\cdot\frac{1}{\sigma(\lambda)}\sim1/\lambda^4\;\;,&\lambda\;\textrm{small}\\
\\ \frac{g\cdot
R^7}{\pi\cdot 48^{1/3}\cdot
G_4\cdot\rho}\cdot\frac{1}{\lambda^8}\cdot\frac{1}{\sigma(\lambda)}\approx
5\cdot 10^{13}\;\;,&\lambda\gsim\;10\;\mu
m\end{array}\right.\label{limit4} \eeq

which is plotted in Fig.~\ref{fig.2} (green dash-dotted line). The
matrix elements above have been calculated using the WKB
approximation for the states. It can be shown, however, that using
the exact Airy functions for the bound states in numerically
calculating the matrix elements produces results, that agree with
the ones derived from the above approximate matrix elements to
within 10\% on the whole range of $\lambda$. One should note that
we have been using here the Yukawa potential $\Delta\phi$
including its constant pieces to yield the full contributions to
$\Delta E_0^{(2)}$ and $\Delta E_0^{(1)}$ (which for $\Delta
E_0^{(1)}$ at large $\lambda$ leads to a behaviour of $\Delta
E_0^{(1)}\sim\lambda^{-2}$ instead of $\lambda^{-1}$, see
eq.~\eqref{shift2} and the one before).

\section{Conclusion}

The recent observation of quantum states of ultra-cold neutrons in
the Earth's gravitational field allows one to derive bounds on the
strength and range of an additional force from the experimental
fact, that the crucial measured parameters of the ground state and the first excited state, their
vertical extensions (which in turn are related to their energy
eigenvalues), have been determined to be consistent with Newtonian
gravity to within a differential positioning uncertainty of
$\delta \Delta h=2.0\,\mu m$ at 90\% confidence level. Such an
analysis is interesting from the point of view of systematic
errors. Since the absence of electric charge and the weakness of
its magnetic moment extremely suppresses the electromagnetic false
effects a neutron is exposed to, bounds on additional force
derived from neutron experiments can be given in a systematically
quite clean way.

The analytical bounds derived here and shown in Fig.~\ref{fig.2} form a consistency check for
numerical analyses like the one of~\cite{ab}. Following from the
turning point shift criterion developed here, our limits are valid
for $\lambda\gsim 1\,\mu m$ because of breakdown of the
perturbation expansion for Yukawa-like forces of smaller range.
Within a range of $\lambda=3\ldots10\,\mu m$ we observe a change
in the negative power of the $\lambda$-dependency of the limit
from -2 towards -1. This invokes the interesting fact that the
expectation of a turning point shift $\delta h\sim\Delta E$
holding for small $\lambda$ fails for longer ranges of the
interaction. For even larger $\lambda$ the presence of the
absorber leads to a $\lambda$-independent upper limit on $\alpha$
of about $2\cdot 10^{11}$ in the attractive case for
$\lambda\gsim 10\,\mu m$, which for this case above
$\lambda\sim10\,mm$ further decreases like $\lambda^{-2}$.

Furthermore, note that the bound at large $\lambda>5\,\mu m$
translates into a bound on the strength of CP-violating
pseudoscalar couplings of the axion within the (previously
experimentally unaccessible) astrophysical axion window which is
$g_p g_s/\hbar c < 2\cdot 10^{-15}$ for $5\,\mu m<\lambda<500\,\mu
m$.

\begin{figure}[ht]
\begin{center}
\vspace{0.4cm}\epsfig{file=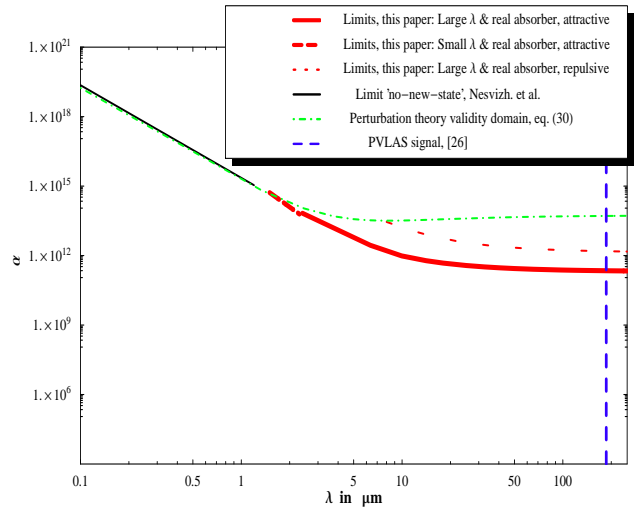,height=2.77in,width=3.44in}
\lbfig{fig.2} \caption{\footnotesize Red thick solid:
\newline Limit for large $\lambda$ with real absorber, attractive case
eq.s~\eqref{limit3} and~\eqref{limit3b}.
\newline Red thick long-dash:
\newline Limit for small $\lambda$ in
presence of a real absorber, attractive case eq.~\eqref{limit5}.
\newline Red thin wide-dash:
\newline Limit for large $\lambda$ with real absorber, repulsive case (eq.~\eqref{limit3}).
\newline Black solid:
\newline The 'no-new-state' limit of
ref.~\cite{nes3}.
\newline Green dash-dot:
\newline Validity boundary of
perturbation theory, eq.~\eqref{limit4}.
\newline Blue vertical dash:
\newline PVLAS signal~\cite{PVLAS}. We display this, as an axion interaction would generate a Yukawa-like interaction with $\alpha\sim g_sg_p$~\cite{wilcz}, and the photon-axion coupling (if PVLAS sees an axion and not a scalar) is - up to a loop suppression - roughly of the same order as the direct axion-nucleon coupling we measure.}
\end{center}
\end{figure}

The analytical limit on $\alpha$ and $\lambda$ for very small
$\lambda<1\,\mu m$ given by the analysis of \cite{nes3} is beyond the scope of our analysis due to its non-perturbative nature. Other
experimental limits on extra forces are derived from mechanical
experiments and can be found, e.g.,
in~\cite{Bordag,ad,Chia,Fischb2,Kapner}. Those
limits are derived from Casimir-force measurements or
mechanical pendulum experiments. They are significantly
better in numbers than the one derived here, however,
one should stress the completely different nature of possible systematical effects present in these micro-mechanical experiments compared to the systematics at work in our
neutron experiment. Casimir-force measurements depend
strongly on the geometry of the experiment and the theoretical treatment of the Casimir effect, which is a difficult task. Therefore, these limits extracted from Casimir-force
measurements are not as rigorous as those shown in Fig.~\ref{fig.2}.
Using an elementary and electrically neutral particle like
the neutron one can provide exclusion limits on additional
interactions untouched by the false effects of the mechanical experiments.

\noindent {\bf Acknowledgements}: We would like to thank
V.~V.~Nesvizhevsky, A. Voronin, K.~V.~Protasov, S.~Nahrwold and C. Krantz for
useful discussions and comments.

This work was supported by the INTAS grant 99-705 and the German
Federal Ministry for Research and Education (BMBF) under contract
number 06 HD 153I. AW received additional support from a research fellowship of INFN (Italy).

\end{document}